%% file: main.tex
\documentclass[10pt,conference]{IEEEtran}
\IEEEoverridecommandlockouts

\usepackage[utf8]{inputenc}
\usepackage[T1]{fontenc}

\usepackage{amsmath,amsthm}
\usepackage[linesnumbered,ruled,vlined]{algorithm2e}
\usepackage{siunitx} 

\usepackage{graphicx}      
\graphicspath{{./images/}} 
\usepackage{caption}       
\usepackage{subcaption}    
\usepackage{tikz}          
\usetikzlibrary{arrows}    

\usepackage{booktabs}
\usepackage{multirow}
\usepackage{balance}
\usepackage{listings}
\usepackage{soul}
\definecolor{hl}{rgb}{0.80,0.80,0.80} 
\sethlcolor{hl}						  
\usepackage{color}
\usepackage{color, colortbl}

\usepackage[hidelinks]{hyperref} 
\usepackage[nameinlink,capitalise]{cleveref} 

\newcommand{\eg}{\textit{e.g.,}}

\newcommand{\ie}{\textit{i.e.,}}

\newcommand{\etal}{\textit{et al.}}

\newcommand{\wrt}{\textit{w.r.t.}}

\newcommand{\evomaster}{\textsc{EvoMaster}}
\newcommand{\mosa}{\textsc{MOSA}}
\newcommand{\mio}{\textsc{MIO}}
\newcommand{\our}{\textsc{LT-MOSA}}
\newcommand{\catwatch}{\textit{CatWatch}}
\newcommand{\features}{\textit{Features-Service}}
\newcommand{\ncs}{\textit{NCS}}
\newcommand{\ocvn}{\textit{OCVN}}
\newcommand{\proxyprint}{\textit{ProxyPrint}}
\newcommand{\scout}{\textit{Scout-API}}
\newcommand{\scs}{\textit{SCS}}

\newcommand{\pvalue}{p\text{-value}}
\newcommand{\atwelve}{\mathrm{\hat{A}}_{12}}

\tikzstyle{mybox} = [draw=black, thick, rectangle, inner ysep=5pt, inner xsep=5pt] 

\def\BibTeX{{\rm B\kern-.05em{\sc i\kern-.025em b}\kern-.08em
    T\kern-.1667em\lower.7ex\hbox{E}\kern-.125emX}}
    
\lstdefinestyle{HTTP}{
    backgroundcolor=\color{white},   
    commentstyle=\color{green},
    keywordstyle=\color{magenta},
    numberstyle=\tiny\color{black},
    stringstyle=\color{purple},
    basicstyle=\ttfamily\small,
    breakatwhitespace=false,         
    breaklines=true,                 
    captionpos=t,                    
    keepspaces=true,                 
    numbers=left,                    
    numbersep=5pt,                  
    showspaces=false,                
    showstringspaces=false,
    showtabs=false,                  
    tabsize=2,
}

\begin{document}

\title{Improving Test Case Generation for REST APIs Through Hierarchical Clustering}

\author{\IEEEauthorblockN{Dimitri Stallenberg}
\IEEEauthorblockA{\textit{Delft University of Technology}\\
Delft, The Netherlands \\
D.M.Stallenberg@tudelft.nl}
\and
\IEEEauthorblockN{Mitchell Olsthoorn}
\IEEEauthorblockA{\textit{Delft University of Technology}\\
Delft, The Netherlands \\
M.J.G.Olsthoorn@tudelft.nl}
\and
\IEEEauthorblockN{Annibale Panichella}
\IEEEauthorblockA{\textit{Delft University of Technology}\\
Delft, The Netherlands \\
A.Panichella@tudelft.nl}
}

\maketitle

\begin{abstract}
  \input{abstract}
\end{abstract}


\input{introduction}
\input{background}
\input{approach}
\input{study}
\input{results}
\input{validity}
\input{conclusion}

\balance

\bibliographystyle{IEEEtran}
\bibliography{references.bib}

\end{document}

%% file: abstract.tex
With the ever-increasing use of web APIs in modern-day applications, it is becoming more important to test the system as a whole. 
In the last decade, tools and approaches have been proposed to automate the creation of system-level test cases for these APIs using evolutionary algorithms (EAs).
One of the limiting factors of EAs is that the genetic operators (crossover and mutation) are fully randomized, potentially breaking promising patterns in the sequences of API requests discovered during the search.
Breaking these patterns has a negative impact on the effectiveness of the test case generation process.
To address this limitation, this paper proposes a new approach that uses Agglomerative Hierarchical Clustering (AHC) to infer a linkage tree model, which captures, replicates, and preserves these patterns in new test cases.
We evaluate our approach, called \our{}, by performing an empirical study on 7 real-world benchmark applications \wrt{} branch coverage and real-fault detection capability. 
We also compare \our{} with the two existing state-of-the-art white-box techniques (\mio{}, \mosa{}) for REST API testing. 
Our results show that \our{} achieves a statistically significant increase in test target coverage (\ie{} lines and branches) compared to \mio{} and \mosa{} in 4 and 5 out of 7 applications, respectively.
Furthermore, \our{} discovers 27 and 18 unique real-faults that are left undetected by \mio{} and \mosa{}, respectively.

%% file: introduction.tex
\section{Introduction}
\label{sec:introduction}


Over the last decade, the software landscape has been characterized by the shift from large monolithic applications to component-based systems, such as microservices.
These systems, together with their many diverse client applications, make heavy use of web APIs for communication.
Web APIs are almost ubiquitous today and rely on well-established communication standards such as SOAP~\cite{curbera2002unraveling} and REST~\cite{fielding2000architectural}.
The shift towards component-based systems makes it ever-increasingly more important to test the system as a whole since many different components have to work together. 
Manually writing system-level test cases is, however, time-consuming and error-prone~\cite{lenarduzzi2020serverless, lenarduzzi2020toward}.


For these reasons, researchers have come up with different techniques to automate the generation of test cases. 
One class of such techniques is search-based software testing. 
Recent advances have shown that search-based approaches can achieve a high code coverage~\cite{campos2018empirical}, also compared to manually-written test
cases~\cite{panichella2017java}, and are able to detect unknown bugs~\cite{fraser20151600,shamshiri2015automatically,arcuri2020handling}.
Search-based test case generation uses Evolutionary Algorithms (EAs) to evolve a pool of test cases through randomized \textit{genetic operators}, namely \textit{mutations} and \textit{crossover/recombination}. 
More precisely, test cases are encoded as chromosomes, while statements (\ie{} method calls) and test data are encoded as the genes~\cite{tonella2004evolutionary}.
In the context of REST API testing, a test case is a sequence of API requests (\ie{} HTTP requests and SQL commands) on specific resources~\cite{arcuri2019restful, arcuri2020handling}.

REpresentational State Transfer (REST) APIs deal with states. 
Each individual request changes the state of the API, and therefore, its execution result depends on the state of the application (\ie{} the previously executed requests). 
\cref{lst:motivation} shows an example of HTTP requests made to a REST API that manages \texttt{products}.
In the example, the first request authenticates the client to the API with the given \texttt{username} and \texttt{password}.
In return, the client receives a \texttt{token} that can be used to make subsequent requests.
The second request creates a new product by specifying the \texttt{id}, \texttt{price}, and the \texttt{token}.
The \texttt{price} is then updated in the third request and the changes are retrieved in the last request.

\begin{lstlisting}[frame=single, label=lst:motivation , caption={Motivating example of patterns in API requests.}, linewidth=0.98\linewidth, xleftmargin=1em, float=t, belowskip=-1.5\baselineskip]
POST   authenticate?user=admin&password=pwd
POST   product?id=1&price=10.99&token={key}
UPDATE product/1?price=8.99&token={key}
GET    product/1?token={key}
\end{lstlisting}

The example above contains patterns of HTTP requests that strongly depend on the previous ones.
The \texttt{GET} request can not retrieve a product that does not exist, and therefore, can not be successfully executed without request 2.
Similarly, the \texttt{UPDATE} request can not be executed before the product is created. Lastly, request 2, 3, and 4, all depend on request 1 for the authentication token.
Hence, HTTP requests should not be executed in any random order~\cite{zhang2019resource}.

Test generation tools rely on EAs to build up sequences of HTTP requests iteratively through genetic operators, \ie{} crossover and mutation~\cite{arcuri2020handling,arcuri2019restful,arcuri2018test}.
While these operators can successfully create promising sequences of HTTP requests, they do not directly recognize and preserve them when creating new test cases~\cite{watson2007building}. For example, the genetic operators may remove request 2 from the test case in \cref{lst:motivation}, breaking requests 3 and 4 unintentionally.

In this paper, we argue that detecting and preserving patterns of HTTP requests, hereafter referred to as \textit{linkage structures}, improves the effectiveness of the test case generation process.
We propose a new approach that uses Agglomerative Hierarchical Clustering (AHC) to infer these linkage structures from automatically generated test cases in the context of REST APIs testing.
In particular, AHC generates a \textit{Linkage Tree} (LT) model from the test cases that are the closest to reach uncovered test targets (\ie{} lines and branches).
This model is used by the genetic operators to determine which sequences of HTTP requests should not be broken up and should be replicated in new tests.

To evaluate the feasibility and effectiveness of our approach, we implemented a prototype based on \evomaster{}, the state-of-the-art test case generation tool for Java-based REST APIs. 
We performed an empirical study with \num{7}~benchmark web/enterprise applications from the \evomaster{} Benchmark (EMB) dataset.
We compared our approach against the two state-of-the-art algorithms for system-level test generations implemented in \evomaster{}, namely Many Independent Objective (MIO) and Many Objective Search Algorithm (MOSA). 



Our results show that \our{} covers significantly more test targets in 4 and 5 out of the 7 applications compared to \mio{} and \mosa{}, respectively.
On average, \our{}, covered \SI{11.7}{\percent} more test targets than \mio{} (with a max improvement of \SI{66.5}{\percent}) and \SI{8.5}{\percent} more than \mosa{} (with a max improvement of \SI{37.5}{\percent}). 
Furthermore, \our{} could detect, on average, \num{27} and \num{18}~unique real-faults that were not detected by \mio{} and \mosa{}, respectively.
 

In summary, we make the following contributions:
\begin{enumerate}
	\item A novel approach that uses Agglomerative Hierarchical Clustering to learn and preserve \textit{linkage structures} embedded in REST API.
	\item An empirical evaluation of the proposed approach with the two state-of-the-art algorithms (\mio{} and \mosa{}) on a benchmark of \num{7} web/enterprise applications.
	\item A full replication package including code and results~\cite{dimitri_stallenberg_2021_5106114}.
\end{enumerate}


The remainder of this paper is organized as follows. \cref{sec:background} summarizes the related work and background concepts. 
\cref{sec:approach} introduces our approach called, \our{}, and gives a detailed breakdown of how it works. 
\cref{sec:study} describes the setup of our empirical study. 
\cref{sec:results} discusses the obtained results and presents our findings. 
\cref{sec:validity} discusses the threats to validity and \cref{sec:conclusion} draws conclusions and identifies possible directions for future work.

%% file: background.tex
\section{Background and Related Work}
\label{sec:background}

This section provides an overview of basic concepts and related work in search-based software testing, REST API testing, test case generation, and linkage learning.

\subsection{Search-based software testing}
\label{sec:search}

Search-based software testing has become a widely used and effective method of automating the generation of test cases and test data~\cite{mcminn2004search, anand2013orchestrated}.
Automatic test case generation significantly reduces the time needed for testing and debugging applications (\eg{}~\cite{soltani2018search}) and it has been successfully used in industry (\eg{} ~\cite{ali2013generating, alshahwan2018deploying}).
Popular tools for automatically generating test cases include EvoSuite~\cite{fraser2011evosuite}, for unit testing, and Sapienz~\cite{alshahwan2018deploying, mao2016sapienz}, for Android testing.

Evolutionary Algorithms (EAs) are one of the most commonly used class of meta-heuristics in search-based testing.
EAs have been used to generate both test data~\cite{mcminn2004search} and test cases~\cite{tonella2004evolutionary}. The latter includes test data, method sequences and assertions.
EAs are inspired by the biological process of evolution.
It initializes and evolves a population of randomly generated individuals (test cases).
These individuals are then evaluated based on a predefined fitness function.
The individuals with the best fitness value are \textit{selected} for reproduction through \textit{crossovers} (swapping elements between two individuals) and random \textit{mutations} (small changes to individuals).
The offspring test cases resulting from the reproduction are then \textit{evaluated}.
Finally, the population for the next generation is created by selecting the best individuals across the previous population and the newly generated tests (\textit{elitism}).
This loop (reproduction, evaluation, and selection) continues until a stopping condition has been reached.
The final test suite is created based on the population's best individuals.

\subsection{REST API Testing}
\label{sec:restful}


A REpresentational State Transfer (REST) API is oriented around resources.
This differs from a command-oriented API like for example the Remote Procedure Call (RPC) standard.
A REST API request performs an action on a specific resource.
These actions are encoded by the different methods defined in the HTTP protocol.
Common HTTP methods include \texttt{GET}, \texttt{HEAD}, \texttt{POST}, \texttt{PUT}, \texttt{PATCH}, and \texttt{DELETE}.
These actions are performed on an endpoint, which is the location of the resource.
An example of this would be performing a \texttt{GET} action on the \fbox{\texttt{/user/3}} endpoint to retrieve the information of a user with a user id of 3.
Another example would be performing a \texttt{POST} action on the \fbox{\texttt{/user/}} endpoint to create a new user.


With the recent rise in popularity of REST APIs in the last decade, it is becoming more important to test this critical communication layer.
There are two different ways system-level API testing can be approached: black-box and white-box testing.
Black-box testing frameworks (\eg{} RESTTESTGEN~\cite{viglianisi2020resttestgen}, EvoMaster black-box~\cite{arcuri2019restful}) examine the functionality of the system without looking at the internals of the system.

In contrast, white-box testing approaches rely on the internal structure of the system and measure the adequacy of the tests based on coverage criteria (\eg{} branch coverage).
This allows the algorithm to easily identify which paths have been covered and which have not.
Prior studies show white-box techniques achieve better results than their black-box counterparts~\cite{arcuri2019restful}.
Additionally, white-box techniques allow to integrate SQL databases in the test case generation process~\cite{arcuri2020handling}.



%

\subsection{Test Case Generation for REST APIs}
\label{sec:evomaster}


\evomaster{} is a tool that aims to generate system-level test cases for REST APIs.
It internally uses evolutionary algorithms to evolve the test cases iteratively.
At the time of writing, \evomaster{} provides two EAs.
The first algorithm is the Many Independent Objective (MIO) algorithm proposed by Arcuri~\etal{}~\cite{arcuri2018test}.
The second algorithm is a variant of the Many-Objective Sorting Algorithm (MOSA) proposed by Panichella~\etal{}~\cite{Panichella:icst2015}.
Both of these algorithms are specifically designed for test case generation and consider the peculiarities of these systems.
The pseudo-code of these algorithms can be found in the respective papers. 

\subsubsection{MIO}
\label{sec:mio}

The Many Independent Objective (MIO) algorithm is an evolutionary algorithm that aims to improve the scalability of many-objective search algorithms for programs with a very large number of testing targets (in the order of millions)~\cite{arcuri2018test}.
It works under the assumption that: (i) each target can be independently optimized; (ii) targets can be strongly related, for example when nested, or completely independent; (iii) not all targets can be covered.
Based on these assumptions, \mio{} maintains a separate population for each target of the System Under Test (SUT).
The fitness function for each population consists solely of the objective of that population.
So, each population compares and ranks the individuals based on a single testing target.
At the start of the algorithm, all populations are empty. 
Each iteration, the algorithm either samples a random new test case with a certain probability or it samples a test case from one of the populations with an uncovered target.
This test case is then added to all populations with an uncovered target and evaluated independently.
The population that is chosen when a test case is sampled from one of the populations, is based on the number of times a test case has been sampled from that population before.
Each time a population is sampled, a counter is incremented. If a test case with a better fitness value is added to the population, the counter is reset to zero.
The sampling mechanism chooses the population with the lowest counter.
This makes sure that the algorithm won't get stuck on an unreachable target.
When a population reaches a certain predefined size, it removes the test case with the worst fitness value.
At the end of the algorithm, a test suite is built with the best test case from each population.

\subsubsection{MOSA}
\label{sec:mosa}

The Many Objective Sorting Algorithm (MOSA) is an evolutionary algorithm that focuses on optimizing multiple objectives (\eg{} branches) at the same time~\cite{Panichella:icst2015}.
It adapts the NSGA-II algorithm, which is one of the most popular multi-objective search algorithms~\cite{deb2002fast}.
In \mosa{}, a test case is represented as a chromosome.
Each testing target (\eg{} branch, line) in the code corresponds to a separate objective measuring the distance (of a given test) toward reaching that target. 
The fitness of the test cases is measured according to a vector of scalar values that represent these different objectives. 
Since \mosa{} has many different objectives, it uses two preference criteria to determine which test cases should be selected and evolved first: (i) minimal distance to the uncovered target; (ii) test case length.
More precisely, it first looks for the subset of the Pareto front that contains test cases with a minimum distance for each uncovered objective.
When multiple test cases are equally close to cover the same target, the smallest test case will be selected. 
In each generation, an archive collects the test cases that cover previously uncovered targets.
The archive is updated every time a newly generated test case covers new targets or covers already covered targets but with fewer statements.

\subsubsection{Comparison}
\label{sec:comparison}

Both \mio{} and \mosa{} produce good results in both unit and system-level tests.
In the context of system-level testing, Arcuri~\cite{arcuri2018test} showed that \mio{} achieves the best average results, but there are web/enterprise applications in which \mosa{} achieves higher coverage. 
In unit testing, Campos~\etal{}~\cite{campos2018empirical} showed that \mosa{} (and its variants) achieves overall better coverage than \mio{}. 
Therefore, in this paper, we consider both \mio{} and \mosa{} as they excel in different scenarios and are the state-of-the-art in test case generation for REST APIs.

Notice that an extension of \mosa{}, called DynaMOSA~\cite{Panichella:tse2018}, has been proposed in the related literature for unit testing.
Compared to \mosa{}, DynaMOSA organizes the coverage targets (\eg{} branches) of a given code unit into a global hierarchy based on their structural dependencies.
Then, the list of search objectives is updated dynamically based on their structural dependencies and the previously covered targets.
While previous studies in unit-testing showed that DynaMOSA outperforms its predecessor \mosa{}~\cite{Panichella:tse2018, campos2018empirical}, it cannot be applied for REST APIs as no global hierarchy exists across the coverage targets of different microservices or functions/classes within the same microservice\footnote{Micro-services are loosely coupled and deployable independently.}.

\subsubsection{Chromosome Representation}
\label{sec:genes}

Test cases in both search algorithms included in \evomaster{} are represented by two genes: action gene and input gene. 
The action gene represents the structure and order of the HTTP requests in the test case. \evomaster{} extracts these actions from the Swagger/OpenAPI documentation that has to be provided for each system under test. An action gene consists of the HTTP method and the REST endpoint. An example of an action gene would be \fbox{\texttt{POST /authentication}}.

The input gene represents the input data for the HTTP request. An example of this input data would be the username and password that are required by the \fbox{\texttt{/authentication}} endpoint. This input data is sampled from the source code of the SUT. 

\subsection{Linkage Learning in EAs}
\label{sec:ml}



\textit{Linkage-learning} refers to a large body of work in the evolutionary computation community that aims to infer \textit{linkage structures} present in promising individuals~\cite{watson1998modeling}. Linkage structures are groups of ``good'' genes that contribute to the fitness of a given population. Accurate inference of \textit{linkage structures} has been used to design ``competent'' genetic operators~\cite{pelikan2001escaping} for numerical problems. These operators are designed to replicate rather than break groups of genes (patterns) into the offspring. 

To learn linkage structures from numerical chromosomes, researchers have used different unsupervised machine learning algorithms.
BOA~\cite{pelikan1999boa} constructs a Bayesian Network and creates new numerical chromosomes using the joint distribution encoded by the network.
DSMGA~\cite{yu2009dependency} uses Dependency Structure Matrix (DSM) clustering and applies the crossover by exchanging gene clusters between parent chromosomes. 
3LO~\cite{przewozniczek} employs local optimization as an alternative method for linkage learning.

Two state-of-the-art EAs for numerical problems are LT-GA~\cite{thierens2010linkage} and GOMEA~\cite{thierens2011optimal}.
Both algorithms use clustering to infer linkage-trees, representing the linkage structures between genes (problem variables) using tree-like structures.
GOMEA uses agglomerative hierarchical clustering as a faster and more efficient way to learn linkage-trees~\cite{thierens2011optimal}.
GOMEA uses the \textit{gene-pool optimal mixing} to create new solutions by applying a local search within the recombination procedure.
More precisely, it creates offspring solutions from one single parent by iteratively replicating (copying) gene clusters from different donors.
In each iteration, the new solution is evaluated; if its fitness improves, the replicated genes are kept; otherwise, the change is reverted.


\textit{Linkage models} have been successfully applied to evolutionary algorithms for numerical~\cite{bouter2017exploiting}, permutation~\cite{bosman2016expanding}, and binary optimization problems~\cite{thierens2010linkage,olsthoorn2021multiobjective} with fixed length chromosomes.
However, test cases for REST APIs are characterized by a more complex structure~\cite{arcuri2019restful}: each test is a sequence of HTTP requests towards a RESTful service, each with input data, such as HTTP headers, URL parameters, and payloads for \texttt{POST}/\texttt{PUT}/\texttt{PATCH} methods. 
Besides, a test case might include SQL data and commands for microservices that use databases~\cite{arcuri2020handling}.
Finally, test cases have a variable size, and their lengths can also vary throughout the generations.
Therefore, we need to tailor existing \textit{linkage learning} methods according to the test case characteristics discussed above.

%% file: approach.tex
\section{Approach}
\label{sec:approach}

This section presents our approach, called \our{}, for system-level test cases generation and that incorporates and tailors linkage learning into \mosa{}~\cite{Panichella:icst2015}.
We selected \mosa{} as the base algorithm to apply linkage learning because it evolves a single population of test cases, which is a requirement for the learning process. 
Additionally, \mosa{} has been proved to be very competitive in the context of RESTful API testing~\cite{arcuri2018test}, unit testing~\cite{panichella2018large, campos2018empirical}, and DNN testing~\cite{haq2020automatic}.

\begin{algorithm}[t]
\footnotesize
\DontPrintSemicolon
\KwIn{\;
 Coverage targets $\Omega=\{\omega_1, \dots, \omega_n\}$\;
 Population size $M$ \;
 Frequency $K$ for updating the linkage tree model \;
 }
\KwResult{A test suite $T$}
\Begin{
$P \longleftarrow$ RANDOM-POPULATION($M$)\;
archive $\longleftarrow$ UPDATE-ARCHIVE($\emptyset$, $P$)\;
Fronts $\longleftarrow $ PREFERENCE-SORTING$(R)$ \;
\While{not (stop\_condition)}
    {
    \textcolor{blue}{
    L $\longleftarrow$ LEARN-LINKAGE-MODEL(Fronts[0], $K$)
    } \;
    $P' \longleftarrow \emptyset$\;  
    \For{index = 1..$M$}{
    	parent $\longleftarrow$ TOURNAMENT-SELECTION($P$)\;
    	
    	\If{apply\_recombination}{
    		\textcolor{blue}{donor $\longleftarrow$ TOURNAMENT-SELECTION($P$)}\;
    		\textcolor{blue}{offspring $\longleftarrow$ LINKAGE-RECOMB(parent, donor, L)}\;
			\textcolor{blue}{offspring $\longleftarrow$ MUTATION(offspring)}\;
    	}
    	\Else{
    		\textcolor{blue}{offspring $\longleftarrow$ MUTATION(parent) }\;
    	}
    	$P' \longleftarrow P' \bigcup$ 
    	$\{$offspring$\}$\;
    	    	archive $\longleftarrow$
UPDATE-ARCHIVE(archive, offspring)\;
    }

    $R \longleftarrow P \bigcup P'$\;
    Fronts $\longleftarrow $ PREFERENCE-SORTING$(R)$ \;
    $P \longleftarrow$ ENVIRONMENTAL-SELECTION(Fronts, $M$)\;

    }
$T \longleftarrow $ archive
}
\caption{\our{}}\label{algo:ltmosa}
\end{algorithm}

Algorithm~\ref{algo:ltmosa} outlines the pseudo-code of \our{}.
The parts where \our{} deviates from \mosa{} are highlighted with a blue color.
\our{} starts with initializing the population $P$ and computing the corresponding objective scores (line 2).
Each test case is composed of HTTP calls (\textit{actions}) and SQL commands (\textit{database actions}) \cite{arcuri2020handling}.
The RANDOM-POPULATION function also executes the generated tests and computes their objective scores using the \textit{branch distance}~\cite{korel1990}.
The \textit{branch distance} is a well-known heuristic in search-based testing to measure how far each test case is from reaching a given coverage target (\eg{} branch).
Then, the test cases are sorted in sub-dominance fronts using the \textit{preference sorting algorithm}~\cite{Panichella:icst2015}, in line 4.
The test cases within the first front (Front[0]) are the closest ones in $P$ to reach the coverage targets and, therefore, the fittest individuals to consider for model learning.

Afterwards, the population $P$ is evolved through subsequent generations within the loop in lines 5-20.
Each generation starts by training a \textit{linkage tree model} on the first non-dominated front (line 6) with the goal of learning patterns of HTTP and SQL actions that strongly contribute to the ``optimality'' of the population.
We discuss the learning procedure in detail in \cref{sec:ltree}.
Once the \textit{linkage tree model} is obtained, \our{} selects the \textit{fittest} test cases using the \textit{tournament selection} (line 9 and 11) and creates an offspring population $P'$ by using a \textit{linkage-based recombination}~\cite{thierens2011optimal} (line 12) and \textit{mutation}~\cite{arcuri2020handling} (line 13 and 15). 
The \textit{linkage-based recombination} is a specialized \textit{crossover} that relies on the \textit{linkage tree model} to decide which patterns of genes (HTTP requests) can be copied into the offspring test cases. 
We describe the \textit{linkage-based recombination} operator in \cref{sec:mixing}.

\our{} adds the newly generated tests into the offspring population (line 16), executes them, and updates the archive (line 17) in case new coverage targets have been reached.
The generation ends by selecting the best $M$ test cases across the existing population $P$ and the offspring population $P'$.
This selection is made by combining the two population into one single pool $R$ of size $2 \times M$ (line 18), applying the \textit{preference sorting} (line 19), and selecting $M$ solutions from the non-dominated fronts starting from Front[0] until reaching the population size $M$ (line 20).

The search stops when the termination criteria are met (condition in line 5), the final test suite will then be composed of all test cases that have been stored in the \textit{archive} throughout the search.
Note that \our{} updates the \textit{archive} in each generation by storing the shortest test case covering each target $\omega_i$.
Finally, the list of objectives is updated such that the search focuses only on the targets (branches) that are left uncovered.

In the following sub-sections, we detail the key novel ingredients in \our{}, namely the \textit{linkage model learning} (LT) (\cref{sec:ltree}), the \textit{linkage-based recombination operator} (\cref{sec:mixing}), and the \textit{mutation} operator (\cref{sec:mutation}).

\subsection{Linkage tree learning}
\label{sec:ltree}
In this section, we describe the main changes we applied to the traditional \textit{linkage learning} and adapt it to our context, \ie{} test case generation for RESTful APIs.

\subsubsection{Linkage Encoding}
\label{sec:encoding}

The first problem we had to solve is encoding the test cases into discrete vectors of equal length, which can be interpreted and analyzed via hierarchical clustering.
To this aim, we opted for \textit{encoding} test cases as binary vectors whose entries denote the presence (or not) of the possible HTTP requests.
Given an SUT (software under tests), there are $N$ possible HTTP requests to the available APIs.
This information can be extracted from the Swagger/OpenAPI definition~\cite{arcuri2019restful}, which is a widely-used tool for REST API documentation.
A Swagger definition contains the HTTP operations available for each API endpoint.
Each operation contains both fixed and variable parts.
The fixed part includes the type of operation (\texttt{POST}, \texttt{GET}, \texttt{PUT}, \texttt{PATCH}, and \texttt{DELETE}
), the IP address or the URL of the target API, and the HTTP headers.
For the variable part, the Swagger definition includes information about the input data (\eg{} \textit{string}, \textit{double}, \textit{date}, etc.) that can vary. 
Therefore, for each API endpoint, we identify the available HTTP operations, hereafter referred  to as \textit{actions}, by parsing the Swagger definition.

Let $\mathcal{S}=\left\{S_1, \dots, S_N\right\}$ be the set of $N$ \textit{HTTP actions} available for the target SUT.
We encode each test case $T$ as a binary string of size $N$ as follows:
\begin{equation}
	E(T) = \langle e_1, \dots, e_N\rangle \;\; with \;\; e_i =
	\left\{
	\begin{array}{lr}
		0 & if \; S_i \notin T\\
		1 & if \; S_i \in T
	\end{array}
	\right.
\end{equation}
In other words, each element $e_i$ in the encoded vector $E(T)$ is set to 1 if the test case $T$ contains the \textit{action} $S_i$; 0 otherwise.
The linkage model is trained on the binary-coded vectors rather than on the original test cases.
This encoding is used to determine, via statistical analysis, which group of  HTTP \textit{actions} often appear together within the fittest test cases, and which ones never occur together.
This information is used to create more efficient \textit{recombination} operators.

\subsubsection{Linkage Model Training}
\label{sec:training}

In this paper, we use \textit{agglomerative hierarchical clustering} (AHC) over other techniques (e.g. Bayesian Networks) for linkage tree learning. This is because prior studies show AHC is more efficient~\cite{thierens2011optimal}.
In particular, we apply the UPGMA (unweighted pair group method with arithmetic mean) algorithm~\cite{Sokal1958ASM}.
In each iteration, UPGMA merges two clusters that are most similar based on the average distance across the data points (genes in our case) in the two clusters.
The similarity between two HTTP actions genes $S_i$ and $S_j$ is computed using the mutual information as suggested by Thierens and Bosman~\cite{thierens2011optimal}:
\begin{equation}
	MI(S_i,S_j) = H(S_i) + H(S_j) - H(S_i, S_j)
\end{equation}
\noindent where $H(.)$ denotes the information entropy~\cite{shannon2001mathematical}. 

Note that \our{} infers the \textit{linkage tree} for the most promising part of the population, \ie{} the first non-dominated front (line 6 in \cref{algo:ltmosa}).
Furthermore, the training process is applied to the encoded test cases according to the schema described in \cref{sec:encoding} rather than on the actual test cases.
Hence, the \textit{linkage tree} obtained with UPGMA captures the hierarchical relationship between HTTP actions in our case.

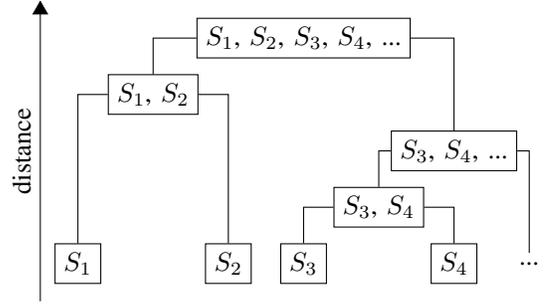
\begin{figure}[t]
\centering
\makeatletter
\begin{tikzpicture}[
sloped, 
squarednode/.style={rectangle, draw=black, fill=white},
]
\node[squarednode] (a) at (-4,0) {$S_1$};
\node[squarednode] (b) at (-2,0) {$S_2$};
\node[squarednode] (c) at (-1,0) {$S_3$};
\node[squarednode] (d) at (1,0) {$S_4$};
\node (e) at (2.0,0) {...};
\node[squarednode] (ab) at (-3,2.25) {$S_1$, $S_2$};
\node[squarednode] (cd) at (0,0.75) {$S_3$, $S_4$};
\node[squarednode] (cde) at (1,1.5) {$S_3$, $S_4$, ...};
\node[squarednode] (all) at (-1,3) {$S_1$, $S_2$, $S_3$, $S_4$, ...};

\draw  (a) |- (ab.west);
\draw  (b) |- (ab.east);
\draw  (c) |- (cd.west);
\draw  (d) |- (cd.east);
\draw  (e) |- (cde.east);
\draw  (cd.north) |- (cde.west);
\draw  (ab.north) |- (all.west);
\draw  (cde.north) |- (all.east);

\draw[->,-triangle 60] (-4.5, -0.5) -- node[above]{distance} (-4.5,3.5);
\end{tikzpicture}
\caption{Example of \textit{linkage tree} model and the Family Of Subset (FOS)}
\label{fig:dendogram}
\end{figure}

For example, let us consider the \textit{linkage tree} depicted in \cref{fig:dendogram}.
In the example, the set of actions $\mathcal{S}=\left\{S_1, S_2, S_3, S_4, \dots\right\}$ (the root of the tree) are partitioned into two clusters: $S_1, S_2$ and $S_3, S_4, \dots$; each sub-cluster can be further divided in sub-cluster until reaching the leaf node.
In general, the linkage tree has $N$ leaves and $N-1$ internal nodes.
The root node contains all HTTP actions of the SUT.
Each internal node divides the set of HTTP actions into two mutually exclusive clusters (the child nodes).
Finally, the leaves contain the individual HTTP actions, which are the starting point of the UPGMA algorithm.

The \textit{linkage tree} nodes are often referred to as
\textit{Family of Subsets} ($\mathcal{F}$) in the related literature~\cite{thierens2010linkage,thierens2011optimal}.
Each node (or subset) $\mathcal{F}' \in \mathcal{F}$ with  $|\mathcal{F}>2|$ has two mutually exclusive subsets (or child nodes) $\mathcal{F}_x$ and $\mathcal{F}_y$ such that $\mathcal{F}_x \bigcap \mathcal{F}_y = \emptyset$ and $\mathcal{F}_x \bigcup \mathcal{F}_y = \mathcal{F}'$. 
Each subset $\mathcal{F}' \in \mathcal{F}$ represents a cluster of HTTP actions that often appear together and characterized the best test cases in the population. 
Therefore, the recombination operator should be applied by preserving these subsets (patterns) when creating new offspring tests.
The next subsection describes the subsets-preserving recombination operator we implemented in \our{}.

The computation complexity of UPGMA is $O(N \times M^2)$, where $N$ is the number of genes and $M$ is the population size.
To reduce its overhead, the linkage tree learning procedure is not applied in each generation.
Instead, the linkage tree model is re-trained every $K$ generations (line 6 of Algorithm~\ref{algo:ltmosa}).

\subsection{Linkage-based Recombination}
\label{sec:mixing}

\mosa{} creates offspring tests using the \textit{single-point crossover}~\cite{Panichella:icst2015}.
This crossover operator is the classic \textit{recombination operator} used in genetic algorithms~\cite{sivanandam2008genetic}, and test case generation~\cite{tonella2004evolutionary, fraser2011evosuite}.
This operator generates two offsprings by randomly swapping statements between two parent tests $T_1$ and $T_2$.
As argued in \cref{sec:introduction}, exchanging statements between test cases in a randomized manner can lead to breaking gene patterns (HTTP actions) that characterized the fittest individuals.
Randomized recombination is also disruptive toward building good partial solutions (building blocks), negatively affecting the overall convergence~\cite{thierens2011optimal}.

Therefore, \our{} uses a \textit{linkage-based recombination operator} rather than the classical \textit{single-point
crossover} to preserve the patterns of HTTP actions identified by the linkage tree model.
The \textit{recombination operator} generates only one offspring starting from two existing test cases, called \textit{parent} and \textit{donor}.
Both test cases are selected from the current population $P$ as indicated in lines 9 and 11.
The offspring is created by copying all genes (HTTP actions with input data) from the \textit{parent} and further injecting only some genes from the \textit{donor}.
These genes are selected by exploiting the \textit{linkage tree model} trained according to Section~\ref{sec:ltree}.

More precisely, we first identify the gene patterns  (\ie{} the subsets $\mathcal{F}' \in \mathcal{F}$) that the \textit{donor} contains.
This is done by iterating across all subsets in the linkage tree model $\mathcal{F}$ and identifying the subsets $\mathcal{F}' \subset \mathcal{F}$ that appear in the encoded vector (see \cref{sec:encoding}) of the \textit{donor}.
\our{} randomly selects one of the identified subsets in $\mathcal{F}'$ and inserts it into the offspring.
The \textit{injection point} is randomly chosen, and the selected genes (HTTP actions with test data) are inserted into the offspring in the exact order as they appear in the \textit{donor}.

If the \textit{donor} does not contain any subset according to the linkage tree (\ie{} $\mathcal{F}'=\emptyset$), then the offspring is generated by applying the traditional \textit{single-point crossover}.
This operator can be applied to the latter case since the \textit{linkage tree model} could not identify any useful gene pattern within the \textit{donor}.


\subsection{Mutation} 
\label{sec:mutation}

In \mosa{}, each test case is mutated with a probability $p_m=1/L$, where $L$ is the test case  length~\cite{Panichella:icst2015}.
This also reflects the existing guidelines in evolutionary computation~\cite{schaffer1989study, smith1996adaptively}, which suggest using a mutation probability $p_m$ proportional to the size of the chromosome.

In recent years, Arcuri~\cite{arcuri2018test} improved the \textit{mutation operator} in the context of system-level test case generation by using a variable mutation rate.
Indeed, the mutation operator in \mio{}~\cite{arcuri2018test} increases the number of mutations applied to each test case from 1 (start of the search) up to 10 (end of the search) with linear incremental steps.
The importance of having a large mutation rate for RESTful API testing has also been confirmed by a recent study  results~\cite{arcuri2020handling}.

Based on these observations, \our{} uses the same mutation rate of \mio{} (\ie{} increasing mutation rate from 1 up to 10 mutations) rather than the fixed mutation rate of \mosa{}.

%% file: study.tex
\section{Empirical Study}
\label{sec:study}

This section details the empirical study that we carried out to evaluate the effectiveness of the proposed solution, called \our{}, and compare it with the state-of-the-art algorithms (\mio{}, \mosa{}) \wrt{} to the following testing criteria: (i) code (line and branch) coverage and (ii) \textit{fault detection} capability.

\subsection{Benchmark}
\label{sec:benchmark}

This study uses the \evomaster{} Benchmark (EMB)\footnote{\url{https://github.com/EMResearch/EMB/releases/tag/v1.0.1}} version 1.0.1.
This benchmark was specifically created as a set of web/enterprise applications for evaluating the test case generation algorithms implemented in \evomaster{}.
We selected this benchmark since it has been widely used in the literature to assess test case generation approaches for REST APIs~\cite{arcuri2019restful, arcuri2018test}.

In this study, we used five real-world open-source Java web applications and two artificial Java web applications. 
\catwatch{} is a metrics dashboard for GitHub organizations. \features{} is a REST Microservice for managing feature models of products.
\ocvn{} (Open Contracting Vietnam) is a visual data analytics platform for the Vietnam public procurement data.
\proxyprint{} is a platform for comparing and making requests to print-shops.
\scout{} is a RESTful web service for the hosted monitoring service ``Scout''.
\ncs{} (Numerical Case Study) is an artificial application containing numerical examples.
\scs{} (String Case Study) is an artificial application containing string manipulation code examples.
We use \ncs{} and \scs{} since they have been designed for assessing test generation tools.
These artificial web applications allow to cover many different scenarios (e.g., deceptive branches~\cite{arcuri2018test}).
Compared to previous studies~\cite{arcuri2019restful, arcuri2018test}, we added the \ocvn{} application as it is the largest real-world system in the benchmark.
We additionally removed the \textit{rest-news} application as it contains artifical examples that are used for classroom teaching.

\cref{tab:benchmark} summarizes the main characteristics of the applications in the benchmark, such as the number of classes, the number of test coverage targets, and the number of endpoints included in the service.
This benchmark contains a total of 1655 classes with around \num{20000} test coverage targets and 440 endpoints, not including tests or third-party libraries.

\evomaster{} requires a test driver for the application under test. This test driver contains a controller that is responsible for starting, resetting, and stopping the SUT. We used the test drivers available in the EMB benchmark for the web applications used in this study.

\begin{table}
\centering
\caption{Web applications from the \evomaster{} Benchmark (EMB) used in the empirical study. Reports the number of Java classes, test coverage targets (\ie{} lines and branches), and the number of endpoints.}
\small
\begin{tabular}{lrrrr}
	\toprule
	Application &  Classes &  Coverage Targets &  Endpoints \\
	\midrule
	CatWatch & 69 & \num{2182} & 23 \\
	Features-Service & 23 & \num{513} & 18 \\
	NCS & 10 & \num{652} & 7 \\
	OCVN & 548 & \num{8010} & 258 \\
	ProxyPrint & 68 & \num{3758} & 74 \\
	Scout-API & 75 & \num{3449} & 49 \\
    SCS & 13 & \num{865} & 11 \\
	\midrule
	Total & \num{1655} & \num{19429} & 440 \\
	\bottomrule
\end{tabular}
\label{tab:benchmark}
\end{table}

\subsection{Research Questions}
\label{sec:questions}

Our empirical evaluation aims to answer the following research questions:

\begin{itemize}
    \item[\textbf{RQ1}] \textit{How does \our{} perform compared to the state-of-the-art approaches with regard to code coverage?}
    \item[\textbf{RQ2}] \textit{How effective is \our{} compared to the state-of-the-art approaches in detecting real-faults?}
    \item[\textbf{RQ3}] \textit{How effective is \our{} at covering test targets over time compared to the state-of-the-art  approaches?}
\end{itemize}

The first two research questions aim to evaluate if preserving patterns in HTTP requests through linkage learning can improve the \textit{effectiveness} of test case generation for REST APIs by reaching a higher coverage and detecting more faults.

The last research question aims to answer if our approach, \our{}, is more \textit{efficient} in covering these test targets by measuring how many test targets are covered at different times within the search budget.

\subsection{Baseline}
\label{sec:baseline}

To answer our research questions, we compare \our{} with the two state-of-the-art search-based test case generation algorithm for REST APIs as a baseline:

\begin{itemize}
    \item \textit{Many Independent Objective (MIO)} is the state-of-the-art for REST API testing, and it is the default search algorithm in \evomaster{}. \mio{} aims to improve the scalability of many-objective search algorithms for programs with a very large number of testing targets (see \cref{sec:mio}).
    \item \textit{Many-Objective Sorting Algorithm (MOSA)} is the base algorithm we use to build and design \our{}. Therefore, we want to assess that our approach outperforms its predecessor. Furthermore, \mosa{} has been proven to be very competitive in the context of REST APIs testing (see \cref{sec:mosa}).
\end{itemize}

\subsection{Prototype Tool}
\label{sec:prototype}


We have implemented \our{} in a prototype tool that extends \evomaster{}, an automated system-level test case generation framework. In particular, we implemented the approach as described in \cref{sec:approach} within \evomaster{}.

The variant of \mosa{} implemented in \evomaster{} differs from the original algorithm proposed by Panichella~\etal{}~\cite{Panichella:icst2015}. The \evomaster{} variant does not use the crossover operator but merely relies on the mutation operator to create new test cases. Therefore, we implemented the \textit{single-point} crossover as described in \cite{Panichella:icst2015} and adapted it to the encoding schema used for representing REST API requests in \evomaster{}.
See \cref{sec:mosa} for more details.

We chose \evomaster{} because it already implements the state-of-the-art test case generation algorithms, and it is publicly available on GitHub. Besides, \evomaster{} implements \textit{testability transformations} to improve the guidance for search-based algorithms~\cite{arcurijuan2020testability} and can handle SQL databases~\cite{arcuri2020handling}.

\subsection{Parameter Setting}
\label{sec:parameters}

For this study, we have chosen to adopt the default search algorithm parameter values set by \evomaster{}.
It has been empirically shown~\cite{arcuri2013parameter} that although parameter tuning has an impact on the effectiveness of a search algorithm, the default values, which are commonly used in literature, provide reasonable and acceptable results.
Thus, this section only lists a few of the most important search parameters and their values:

\subsubsection{Search Budget}
We chose a search budget (stopping condition) based on time instead of the number of executed tests. This choice was made as search time provides the fairest comparison given that we consider different kinds of algorithms with diverse internal routines (also in terms of computational complexity).
Additionally, practitioners will often only allocate a certain amount of time for the algorithm to run.
The search budget for all algorithms was set to \num{30}~minutes as this strikes a balance between giving the algorithms enough time to explore the search space and making the study infeasible to execute.
If the algorithm has covered all its test objectives, it will stop prematurely. Note that running time is considered a less biased stopping condition than counting the number of executed tests since not all tests have the same running time~\cite{panichella2017java, arcuri2020handling, arcuri2018test, fraser2011evosuite}. We further discuss this aspect in the threats to validity.

\subsubsection{\mio{} parameters}
For \mio{}, we used the default settings as provided in the original paper by Arcuri~\etal{}~\cite{Arcuri2017, arcuri2018test}.

\begin{itemize}
	\item \textit{Population size:} We use the default population size of 10 individuals per testing target. Notice that \mio{} uses separate populations for the different targets.
	\item \textit{Mutation:} We use the default number of applied mutations on sampled individuals, which linearly increases from 1 to 10 by the end of the search.
	\item \textit{F:} We use the default percentage of time after which a focused search should start of 0.5.
	\item \textit{$P_r$:} We use the default probability of sampling at random, instead of sampling from one of the populations, of 0.5. This value will linearly increase/decrease based on the consumed search budget and the value of $F$.
\end{itemize}

\subsubsection{\mosa{} parameters}
For \mosa{}, we used the default settings described in the original paper~\etal{}~\cite{Panichella:tse2018}.

\begin{itemize}
	\item \textit{Population size:} 50 individuals (test cases).
	\item \textit{Mutation:} We use the \textit{uniform mutation}, which either changes the test case structures (adding, deleting, or replacing API requests) or the input data. Test structure and test data mutation are equally probable, i.e, each has 50\% probability of being applied. The mutation probability for each statement/data gene is equal to 1/$n$, where $n$ is the number of statements in the test case.
	\item \textit{Recombination Operator:} We use the single-point crossover with a crossover probability of 0.75.
	\item \textit{Selection:} We use the tournament selection with the default tournament size of 10.
\end{itemize}

\subsubsection{\our{} parameters} For \our{}, we used the same parameters as for the \mosa{} algorithm  except for the mutation operator, for which we use the mutation described in Section~\ref{sec:mutation}. Additionally, we use the following parameter values for the \textit{linkage learning} model:

\begin{itemize}
	\item \textit{Frequency:} We use a frequency of 10 generations for generating a new Linkage-Tree model. From a preliminary experiment that we have performed, this provides a balance between having too much overhead ($< 10$) and having an outdated model ($> 10$).
	\item \textit{Recombination Operator:} We use the linkage-based recombination with a probability of 0.75.
\end{itemize}

\subsection{Real-fault Detection}
\label{sec:faults}

To find out the number of unique faults that the search algorithms can detect, \evomaster{} checks the returned status codes from the HTTP requests for \texttt{5xx} server errors, as an indicator for a fault.
Since web applications handle many different clients, when an error occurs it is not desirable for the application to crash or exit as this would also impact the other clients. Thus, web applications return a status code in the \texttt{5xx} range, indicating an error has occurred on the server's side. 
\evomaster{} keeps track of the last executed statement in the SUT (excluding third-party libraries) when a \texttt{5xx} status code is returned, to distinguish between different errors that happen on the same endpoint.

\subsection{Experimental Protocol}
\label{sec:protocol}

For each web application, all three search algorithms (\mosa{}, \mio{}, \our{}) are separately executed, and the resulting number of test targets that are covered is recorded.

Since all three search algorithm used in the study are randomized, we can expect a fair amount of variation in the results.
To mitigate this, we repeated every experiment \num{20}~times, with a different random seed, and computed the average (median) results.
In total, we performed \num{420}~executions, three search algorithms for seven web applications with 20 repetitions each.
With each execution taking 30 minutes, the total execution time is \num{8.75}~days of consecutive running time.

To determine if the results (\ie{} code coverage and fault detection capability) of the three different algorithms are statistically significant, we use the unpaired Wilcoxon rank-sum test~\cite{conover1998practical} with a threshold of \num{0.05}.
This is a non-parametric statistical test that determines if two data distributions are significantly different.
Since we have three different data distributions, one for each search algorithm, we perform the Wilcoxon test pairwise between each configuration pair: (i) \our{} and \mosa{}; (ii) \our{} and \mio{}.
We combine this with the Vargha-Delaney statistic~\cite{vargha2000critique} to measure the effect size of the result, which determines how large the difference between the two configuration pairs is.

To determine how the two configuration pairs compare in terms of efficiency, we analyze the code coverage at different points in time.
While the effectiveness measures the code coverage only at the end of the allocated time, we also want to analyze how algorithms perform during the search.
One way to quantify the efficiency of an algorithm is by plotting the number of test targets at predefined intervals during the search process.
This is called a convergence graph.
We collected the number of targets that have been covered for every generation of each independent run.
To express the efficiency of the experimented algorithms using a single scalar value, we computed the overall convergence rate as the Area Under the Curve (AUC) delimited by the convergence graph. This metric is normalized by dividing the AUC in each run by the maximum possible AUC per application\footnote{Which corresponds to the area of the box with a height of the maximum code coverage and a width equal to the search budget.}.

%% file: results.tex
\section{Results}
\label{sec:results}

This section details the results of the empirical study with the aim of answering our research questions.

\subsection{RQ1: Code Coverage}

\begin{table}[t]
\centering
\caption{Median number of covered test targets.}
\resizebox{0.49\textwidth}{!}{
\begin{tabular}{l rr | rr | rr}
	\toprule
	\multirow{2}{*}{Application} & \multicolumn{2}{c}{\mio{}} & \multicolumn{2}{c}{\mosa{}} & \multicolumn{2}{c}{\our{}} \\
	\cmidrule{2-7} & Median & IQR & Median & IQR & Median & IQR \\
	\midrule
	CatWatch & 1173.00  & 12.75 & 1177.00 & 132.00 & 1215.50 & 161.75 \\
	Features-Service & 488.00 & 72.25 & 455.50 & 33.25 & 478.00 & 5.00\\
	NCS & 622.50 & 1.25 & 623.00 & 4.00 & 622.00 & 3.25 \\
	OCVN & 2421.50 & 374.75 & 2931.50 & 271.00 & 4031.50 & 338.75 \\
	ProxyPrint & 1485.50 & 16.25 & 1501.00 & 78.25 & 1602.50 & 59.00 \\
	Scout-API & 1727.50 & 54.75 & 1707.00 & 69.00 & 1826.50 & 33.25 \\
	SCS &  853.00 & 5.50 & 852.00 & 8.00 & 853.00 & 3.00 \\
	\bottomrule
\end{tabular}
}
\label{tab:coverage-raw}
\end{table}

\cref{tab:coverage-raw} reports the median and inter-quartile range (IQR) of the number of test targets covered by \mio{}, \mosa{}, and \our{} for each of the seven applications.

From \cref{tab:coverage-raw}, we observe that \our{} achieved the highest median value (avg. +334.75 targets) for four out of the seven applications, and \mosa{} and \mio{} both achieved the highest median value (+10.00 and +0.5 targets, respectively) for 1 out of the 7 applications.
The largest increase in code coverage is observable for \ocvn{}, for which \our{} covered +1100.00 more targets.
For \scs{}, both \our{} and \mio{} covered the same number of targets (853.00).
For both artificial applications, namely \ncs{} and \scs{}, the difference between the search algorithms is minimal ($\leq 1$).

In terms of variability (IQR), there is no clear trend with regard to the applications under test and/or the search approaches.
For example, in some cases, the winning configuration (\our{} on \catwatch{}) has the highest IQR with a significant margin (161.75 vs. 132.00 or 12.75). On \scout{}, \our{} yields the lowest IQR by a significant margin (33.25 vs. 8.00 or 5.50).
Within and across each search algorithm, the IQR varies. 

\begin{table}[t]
\centering
\caption{Statistical results ($\pvalue$ and $\atwelve{}$) for the covered test targets (\textit{RQ1}). Significant $\pvalue{}$s (\ie{} $\pvalue < 0.05$) are marked gray.}
\resizebox{0.48\textwidth}{!}{
\begin{tabular}{l rl | rl}
	\toprule
	\multirow{2}{*}{Application} & \multicolumn{2}{c}{\our{} vs \mio{}} & \multicolumn{2}{c}{\our{}  vs \mosa{}} \\
	\cmidrule{2-5} & $\pvalue$ & $\atwelve$ & $\pvalue$ & $\atwelve$  \\
	\midrule
	CatWatch &  \hl{$<$0.01} & 0.87 (Large) & \hl{0.04} & 0.66 (Small) \\
	Features-Service &  0.34 & 0.54 & \hl{$<$0.01} & 0.83  (Large)\\
	NCS &  0.86 & 0.49 & 0.90 & 0.38 (Small) \\
	OCVN &  \hl{$<$0.01} & 1.00 (Large) & \hl{$<$0.01} & 1.00 (Large) \\
	ProxyPrint & \hl{$<$0.01} & 1.00 (Large) & \hl{$<$0.01} & 0.86 (Large) \\
	Scout-API & \hl{$<$0.01} & 0.96 (Large) & \hl{$<$0.01} & 0.96 (Large) \\
	SCS & 0.86 & 0.40 (Small) & 0.50 & 0.50 \\
	\bottomrule
\end{tabular}
}
\label{tab:coverage-statistics}
\end{table}

\cref{tab:coverage-statistics} reports the statistical significance ($\pvalue$), calculated by the Wilcoxon test, of the difference between the number of targets covered by \our{} and the two baselines, \mio{} and \mosa{}.
It also reports the magnitude of the differences according to the Vargha-Delaney $\atwelve$ statistic.

From \cref{tab:coverage-statistics}, we can observe that for the non-artificial web applications, \our{} achieves a significantly higher code coverage than \mio{} in four out of five applications with a \textit{large} effect size ($\atwelve$ statistics). \our{} significantly outperforms \mosa{} in all five applications. The effect size is
 \textit{large} in four applications and \textit{small} for \catwatch{}. 
For the two artificial applications, \ncs{} and \scs{}, there is no statistical difference between the results of \our{} and the two baselines (\mio{} and \mosa{}).
This confirms our preliminary results reported in \cref{tab:coverage-raw}. Moreover, the difference between 
\our{} and \mio{} is not significant for \features{}. Finally, in none of the applications in our benchmark, neither of the baselines achieved a significantly larger coverage than \our{}.

\vspace*{2mm}
\hspace*{-5mm}
\begin{tikzpicture}
\node [mybox] (box){%
\centering
\begin{minipage}{.465\textwidth}
In summary, \our\ achieves significantly higher (most of the cases) or equal code coverage when applied to REST APIs as compared to both \mio{} and \mosa{}.
\end{minipage}
};
\end{tikzpicture}%

\subsection{RQ2: Fault Detection Capability}

\begin{table}[t]
\centering
\caption{Median number of detected real-faults.}
\resizebox{0.48\textwidth}{!}{
\begin{tabular}{l rr | rr | rr}
	\toprule
	\multirow{2}{*}{Application} &  \multicolumn{2}{c}{\mio{}} &  \multicolumn{2}{c}{\mosa{}} &  \multicolumn{2}{c}{\our{}} \\
	\cmidrule{2-7} & Median & IQR & Median & IQR & Median & IQR \\
	\midrule
	CatWatch & 13.00 & 0.25 & 12.00 & 2.00 & 13.50 & 2.25 \\
	Features-Service & 17.00 & 0.00 & 17.00 & 0.00 & 18.00 &  0.50 \\
	NCS &  0 & 0.00 & 0 & 0.00 & 0 & 0.00 \\	
	OCVN & 34.00 & 5.25 & 37.50 & 5.25 & 48.00 & 3.50 \\
	ProxyPrint & 32.50 & 1.00 & 33.00 & 1.00 & 34.00 & 0.25 \\
	Scout-API & 54.50 & 3.75 & 60.00 & 1.50 & 64.00 & 3.00 \\
	SCS & 0 & 0.00 & 0 & 0.00 & 0 & 0.00\\
	\bottomrule
\end{tabular}
}
\label{tab:faults-raw}
\end{table}

\cref{tab:faults-raw} reports the median number of real-faults (and the corresponding IQR) detected by \mio{}, \mosa{}, and \our{} for each of the seven applications.

We observe that for both the artificial applications, \ncs{} and \scs{}, the number of faults that have been detected by any search algorithm is zero. This is because these artificial applications are not designed to fail softly by returning \texttt{5xx} faults.
For the open-source applications, \our{} detects the largest number of faults (avg. +3.40 faults) in all five cases.
The largest increase in fault-detection rate is observable for the \ocvn{} application, with +10.5 more faults detected by \our{} than the baselines.
It is noteworthy that the largest difference between \our{} and the baselines is on the \ocvn{} application, which is the application with by far the most classes (\ie{} 548) and endpoints (\ie{} 258) in our benchmark.
This could be explained by the fact that \our{} also achieved a much higher code coverage for this application.
However, the difference in detected faults for \ocvn{} is larger than for the other applications in the benchmark, which could indicate that \our{} is especially effective for testing large REST APIs.
The faults detected by \our{} are a superset of the faults detected by \mio{} and \mosa{}.
These newly discovered faults originate from the additional coverage that \our{} achieves.

\begin{table}[t]
\centering
\caption{Statistical results ($\pvalue$ and $\atwelve{}$) for the detected real-faults (\textit{RQ2}). Significant $\pvalue{}$s (\ie{} $\pvalue < 0.05$) are marked gray.}
\resizebox{0.48\textwidth}{!}{
\begin{tabular}{l rl | rl}
	\toprule
	\multirow{2}{*}{Application} & \multicolumn{2}{c}{\our{} vs \mio{}} & \multicolumn{2}{c}{\our{} vs \mosa{}} \\
	\cmidrule{2-5} & $\pvalue{}$ & $\atwelve$ & $\pvalue$ & $\atwelve$ \\
	\midrule
	CatWatch &  \hl{$<$0.01} & 0.7 (Large) & \hl{$<$0.01} & 0.84 (Large) \\
	Features-Service & \hl{$<$0.01} & 0.92 (Large) & \hl{$<$0.01} & 0.98 (Large) \\
	NCS &  - & - & - & - \\	
	OCVN & \hl{$<$0.01} & 0.99 (Large) & \hl{$<$0.01} & 0.92 (Large) \\
	ProxyPrint & \hl{$<$0.01} & 0.89 (Large) & \hl{0.03} & 0.66 (Small) \\
	Scout-API & \hl{$<$0.01} & 1.00 (Large) & \hl{$<$0.01} & 0.91 (Large) \\
	SCS &  - & - & - & - \\
	\bottomrule
\end{tabular}
}
\label{tab:faults-statistics}
\end{table}

\cref{tab:faults-statistics} reports the results of the statistical test, namely the Wilcoxon test, applied to the number of faults detected by \our{} and the two baselines, \mio{} and \mosa{}.
It also reports the magnitude of the differences (if any) obtained with the Vargha-Delaney $\atwelve$ statistic.
Significant $\pvalue$s (\ie{} $\pvalue < 0.05$) are highlighted with gray color.
From \cref{tab:faults-statistics}, we can observe that \our{} detects a significantly higher number of faults than \mio{} and \mosa{} in all non-artificial applications.
The effect size ($\atwelve$) is \textit{large} in all comparisons, except for \proxyprint{}, where the effect size is \textit{small} when comparing \our{} and \mosa{}.
Since none of the algorithms detected any faults in the artificial applications, \cref{tab:faults-statistics} does not report any $\pvalue$ or $\atwelve$ statistics for these applications. 

\vspace*{2mm}
\hspace*{-5mm}
\begin{tikzpicture}
\node [mybox] (box){%
\centering
\begin{minipage}{.465\textwidth}
In summary, we can conclude that \our{} detects more faults than the state-of-the-art approaches, namely \mio{} and \mosa{}, for all applications in our benchmark.
\end{minipage}
};
\end{tikzpicture}%

\subsection{RQ3: Code Coverage over Time}

\begin{table}[t]
\centering
\caption{Median normalized AUC for the number of covered test targets. The highest values are marked in gray.}
\begin{tabular}{l c | c | c}
	\toprule
	Application &  \mio{} & \mosa{} & \our{} \\
	\midrule
	CatWatch & 0.77 & 0.78 & 0.78 \\
	Features-Service & 0.78 & 0.75 & \hl{0.82} \\
	NCS & 0.99 & 0.99 & 0.99 \\	
	OCVN & 0.50 & 0.50 & \hl{0.66} \\
	ProxyPrint & 0.87 & 0.82 & \hl{0.88} \\
	Scout-API & 0.84 & 0.81 & \hl{0.86} \\
	SCS & 0.95 & 0.96 & 0.96 \\
	\bottomrule
\end{tabular}
\label{tab:auc-raw}
\end{table}


\begin{figure}[t]
\centering
\includegraphics[width=0.9\columnwidth]{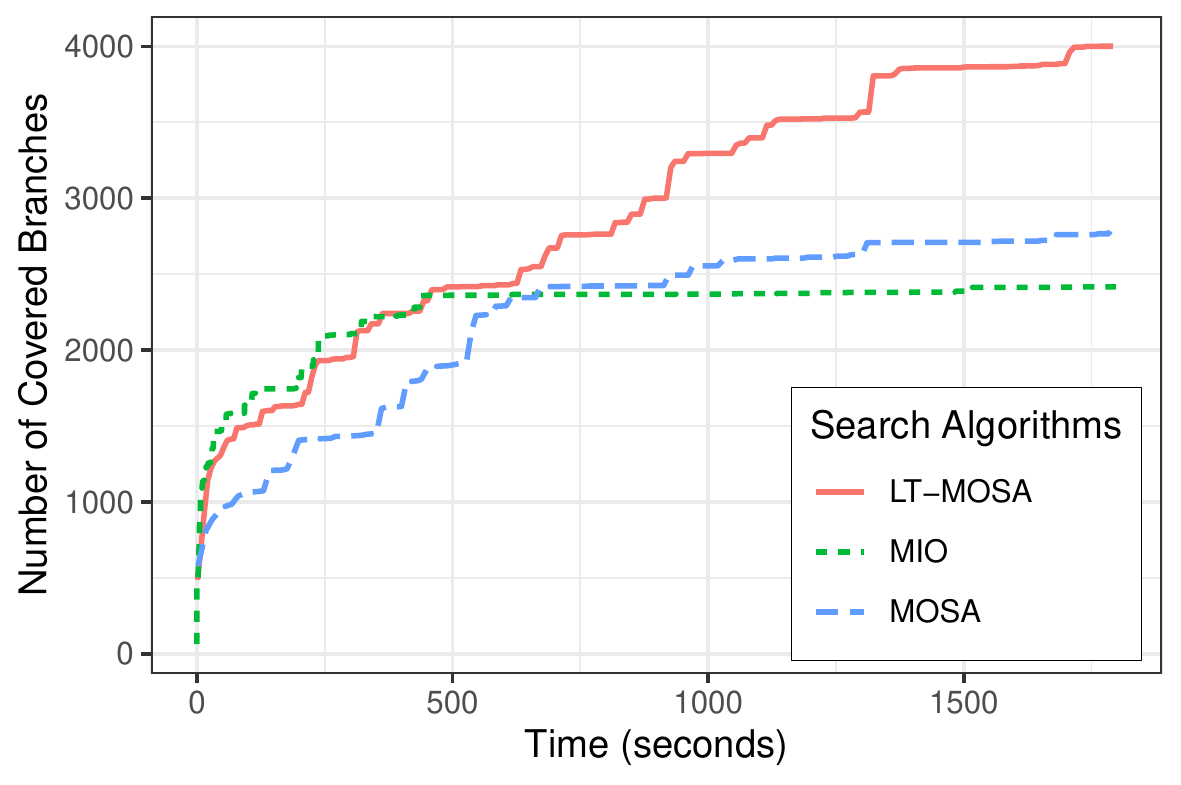}
\caption{Average number of targets covered by our approach (\our{}) and the baselines (\mosa{}, \mio{}) for \ocvn{}.}
\label{fig:branches-ocvn}	
\end{figure}

\cref{tab:auc-raw} reports the median Area Under the Curve (AUC) related to the number of targets covered over time by \mio{}, \mosa{}, and \our{} for each of the seven applications. 
The AUC indicates how efficient the search algorithms are at reaching a certain code coverage.
For more information on how the AUC is calculated and normalized see \cref{sec:protocol}.
\cref{tab:auc-raw} highlights the search algorithm (in gray color) that achieved the highest AUC value.

We observe that for the open-source applications, \our{} has the highest AUC (avg. +0.06) in four out of five applications, with the largest difference (+0.16) in the \ocvn{} application.
For \catwatch{}, both \mosa{} and \our{} have the same AUC (\ie{} 0.78). From \cref{tab:coverage-raw,tab:coverage-statistics} however, we can see that \our{} covers significantly more targets (+38.5) after 30 minutes of search budget.
This means that \mosa{} reaches a higher coverage in the beginning but loses to \our{} over time.


\cref{fig:branches-ocvn} shows the (median) number of targets covered over time by the different search algorithms for \ocvn{}, which is the largest application in our benchmark.
In the beginning of the experiment (0-500 seconds), \mio{} and \our{} perform roughly equal. After the first 500 seconds, \our{} outperforms \mio{}.
This results in a much larger AUC value (+ 0.16) for \our{} compared to \mio{} as indicated in \cref{tab:auc-raw}.
We conclude that \our{} significantly outperforms both \mosa{} and \mio{} in term of effectiveness and efficiency on this application.
To reaffirm this, we can observe that in \cref{fig:branches-ocvn}, \mio{} never reaches 2700 covered targets, \mosa{} takes 1311 seconds to reach that many targets, and \our{} performs this in just 713 seconds, almost half the time of \mosa{}.

For the two artificial applications, \ncs{} and \scs{}, the difference in AUC between the three search algorithms is very minimal ($\leq 0.01$).
From \cref{tab:coverage-raw}, we can also see that \our{} covers one target more than \mosa{} on \scs{} and one target less on \ncs{}.
However, they both yield the same AUC, \ie{} 0.96 (\scs{})  and 0.99 (\ncs{}).
These results are in line with the results from \textit{RQ1}.

\vspace*{2mm}
\hspace*{-5mm}
\begin{tikzpicture}
\node [mybox] (box){%
\centering
\begin{minipage}{.465\textwidth}
In summary, we can conclude that \our{} achieves higher AUC values than the baselines, i.e., it covers more targets and in less time.
\end{minipage}
};
\end{tikzpicture}%

%% file: validity.tex
\section{Threats to Validity}
\label{sec:validity}

This section discusses the potential threats to the validity of the study performed in this paper.

Threats to \textit{construct validity}. We rely on well-established metrics in software testing to compare the different test case generation approaches, namely code coverage, fault detection capability, and running time. As a stopping condition for the search, we measured the search budget in terms of running time (\ie{} 30 minutes) rather than considering the number of executed tests, or HTTP requests. Given that the different algorithms in the comparison use different genetic operators, with different overhead, execution time provides a fairer measure of time allocation.




Threats to \textit{external validity}. An important threat regards the number of web services in our benchmark.
We selected seven web/enterprise applications from the EMB benchmark.
The benchmark has been widely used in the related literature on testing for REST APIs. 
The applications are diverse in terms of size, application domain, and purpose.
Further experiments on a larger set of web/enterprise applications would increase the confidence in the generalizability of our study. 
A larger empirical evaluation is part of our future agenda.


Threats to \textit{conclusion validity} are related to the  randomized nature of EAs. To minimize this risk, we have performed each experiment 20 times with different random seeds. We have followed the best practices for running experiments with randomized algorithms as laid out in well-established guidelines~\cite{arcuri2014hitchhiker}  and analyzed the possible impact of different random seeds on our results. We used the unpaired Wilcoxon rank-sum test and the Vargha-Delaney $\atwelve{}$ effect size to assess the significance and magnitude of our results.

%% file: conclusion.tex
\section{Conclusions and Future Work}
\label{sec:conclusion}


In this paper, we have used agglomerative hierarchical clustering to learn a \textit{linkage tree model} that captures promising patterns of HTTP requests in automatically generated system-level test cases. We proposed a novel algorithm, called \our{}, that extends state-of-the-art approaches by tailoring and incorporating \textit{linkage learning} within its genetic operators. Linkage learning helps to preserve and replicate patterns of API requests that depend on each other. 

We implemented \our{}, in \evomaster{} and evaluated it on seven web applications from the EMB benchmark. Our results show that \our{} significantly improves code coverage and can detect more faults than two state-of-the-art approaches in REST API testing, namely \mio{}~\cite{arcuri2018test} and \mosa{}~\cite{Panichella:icst2015}. This suggests that using unsupervised machine learning (and agglomerative hierarchical clustering in our case) is a very promising research direction.

Based on our promising results, there are multiple
potential directions for future works. In this paper, we used the UPGMA algorithm for hierarchical clustering. Therefore, we intend to investigate more learning algorithms within the hierarchical clustering category. We also plan to investigate other categories of machine learning methods alternative to hierarchical clustering, such as Bayesian Network~\cite{pelikan1999boa}. Finally, \our{} uses a fixed parameter $K$ for the linkage learning frequency. We plan to investigate alternative, more adaptive mechanisms to decide whether the linkage tree model needs to be retrained or not. Finally, we intend to implement and apply linkage learning to unit-test case generation as well.

